\documentclass[letterpaper,superscriptaddress,english,aps,prb,twocolumn,floatfix, showpacs, amsfonts, amssymb]{revtex4}

\usepackage{epsfig, graphicx,psfrag,amsmath,amssymb,float}
\usepackage[T1]{fontenc}
\usepackage[latin9]{inputenc}
\usepackage{amsmath}
\usepackage{amssymb}

\usepackage{amscd}
\usepackage{bm}
\usepackage{psfrag}

\usepackage{babel}

\newcommand{\be}{\begin{equation}}
\newcommand{\ee}{\end{equation}}
\newcommand{\bea}{\begin{eqnarray}}
\newcommand{\eea}{\end{eqnarray}}

\newcommand{\mc}{\mathcal}

\begin{document}
\title{Orbital multicriticality in spin gapped quasi-1D antiferromagnets}

\author{Eran Sela}
\affiliation{Institute for Theoretical Physics, University of Cologne, 50937 Cologne, Germany}
\author{R. G. Pereira}
\affiliation{Instituto de F\'{\i}sica de S\~{a}o Carlos, Universidade de S\~{a}o Paulo, C.P. 369, S\~{a}o Carlos, SP 13560-970, Brazil}

\begin{abstract}
Motivated by the quasi-1D antiferromagnet CaV$_2$O$_4$, we explore spin-orbital systems in which the spin modes are gapped but orbitals are near a macroscopically degenerate classical transition. Within a simplified model we show that gapless orbital liquid phases possessing power-law correlations may occur without the   strict condition of a continuous orbital symmetry. For the model  proposed for CaV$_2$O$_4$, we find that an orbital phase with coexisting order parameters emerges from a multicritical point.
The effective orbital model consists of zigzag-coupled transverse field Ising chains. The corresponding global phase diagram is constructed using field theory methods and analyzed near the multicritical point with the aid of an exact solution of a zigzag XXZ model.
\end{abstract}


\maketitle

\section{Introduction}
\label{se:int}
The orbital degree of freedom plays a major role in the physical properties of strongly correlated transition metal oxides.\cite{tokura} For Mott insulators in which the  low energy electrons occupy orbitals with $e_g$ symmetry, orbital order typically  occurs due to Jahn-Teller  distortions at relatively high temperatures and determines the spatial anisotropy of spin exchange interactions at low temperatures.\cite{kugel}  However, in $t_{2g}$ systems the Jahn-Teller interactions are much weaker and the orbitals can remain unquenched down to low temperatures.\cite{khaliullin} The stabilization of orbital phases may eventually depend on the coupling between spins and orbitals. This coupling can stem either from  the on-site relativistic term or from Kugel-Khomskii~\cite{kugel} type coupling. Particularly intriguing is the possibility of quantum melting of orbital order. One-dimensional (1D) models with SU(2) symmetry for both $S=1/2$ spins and $\tau=1/2$ orbital pseudospins  are known to host liquid phases.\cite{itoi} On the other hand, the 1D model with $S=1$ spins accounts for the orbital Peierls state observed in YVO$_3$.\cite{ulrich} In three dimensions
 an orbital liquid (OL) was proposed as the ground state of LaTiO$_3$,~\cite{maekawa} but was later dismissed by experiments.\cite{hemberger}

In fact, it may be even possible to invert the usual order and freeze out the spins at higher energies than the orbitals. Here we will show that this can be achieved in 1D Haldane gapped spin systems    in the vicinity of a classical orbital transition.
Our theory is motivated by experiments on the $S=1$ zigzag compound CaV$_2$O$_4$.\cite{niazi,pieper} In this material the twofold single-ion orbital degeneracy is lifted by a structural transition below $T_S \approx 141$ K, whereas the strongest intrachain exchange coupling is estimated around $J\approx 230$ K. Antiferromagnetic order only sets in at $T_N \approx 71$ K. Experiments favor a scenario of orbital ordering that turns the  $S=1$ zigzag chains into  $S=1$ ladders.\cite{pieper} Recently, Chern and Perkins \cite{chernperkins}  introduced a model for CaV$_2$O$_4$ which contains Ising  antiferro-orbital (AFO) interactions. These interactions stabilize a  classical AFO ordered phase at weak structural distortion (small orbital energy splitting), whereas a para-orbital (PO) phase corresponding to the experimentally favored ladder structure occurs at strong distortion. Their model at zero structural distortion already presents rich physics occurring as function of increasing relativistic on-site spin-orbit coupling: Chern {\it et al.}\cite{Chern10} found that first an Ising transition turns the spin $S=1$ system from the Haldane spin-liquid to a N\'eel phase, across which the orbitals maintain their finite Ising order, which however is destroyed at a second Ising transition at yet stronger spin-orbit coupling (they have considered ferro-orbital rather than AFO interactions, which however leads to similar physics). More recently, Nersesyan {\it et al.}\cite{nersesyan} found  that whereas the above two-stage ordering scenario occurs for the limiting case of large orbital gap as compared to the spin gap, in the opposite limiting case of large spin gap a single Gaussian transition  between the Haldane spin liquid  and the N\'eel spin-orbital ordered phase takes place as a function of the strength of the spin-orbit coupling. In close connection, spin-orbit interaction   appears to be important for the still unclear orbital order in related spinel compounds.\cite{tsunetsugu,tchernyshyov,wheeler}

In this work we consider the effects of weak spin-orbit interactions near a macroscopically degenerate classical orbital transition, such as the AFO to PO  transition which occurs in the model proposed  for CaV$_2$O$_4$.\cite{chernperkins} We show that this degeneracy renders the orbitals susceptible to arbitrarily small quantum fluctuations induced by coupling to gapped spin modes. As a result, the classical transition point develops into a \emph{multicritical point} (MCP). We describe the phases that appear near the MCP and the universality classes of the transitions.

The outline of the paper is as follows. In Sec.~\ref{se:toy} we study a simplified model in which an Ising orbital chain and a Haldane spin chain are coupled  only by  relativistic spin-orbit interaction. We show that by integrating out the spins one obtains an effective orbital model whose phase diagram contains a MCP, with two   intermediate phases induced by the spin-orbit coupling. One of these phases is a critical orbital liquid,  described as a Luttinger liquid (LL) with power-law correlations, and forms without the strict condition of  continuous orbital symmetry.  In Sec.~\ref{se:Zigzag} we study the model proposed to describe CaV$_2$O$_4$, which besides relativistic spin-orbit coupling  contains Kugel-Khomskii type interactions. We consider particularly the experimentally relevant regime in which the next-nearest-neighbor exchange  coupling in the zigzag structure is much stronger than the nearest-neighbour coupling.  In order to  construct  the phase diagram of this zigzag model, we first present an exact solution of a U(1)$\otimes$U(1) symmetric  model of hard core particles in Sec.~\ref{se:Exact}. We find that the phase diagram of the latter also contains a MCP, but with a single intermediate gapless phase interpreted as an orbital Luttinger liquid that breaks the symmetry between the legs of the zigzag chain. We then return to the model for CaV$_2$O$_4$  and show that in this case the orbital Luttinger liquid  is unstable and gets replaced by a gapped phase with coexisting order parameters.  The existence of this phase in the limit of weak spin-orbit coupling is supported by a global analysis of the phase diagram based on field theory methods carried out in Sec.~\ref{se:Global}.  Finally, the conclusions are presented in Sec. \ref{sec:concl}.

\section{Toy model}
\label{se:toy}
We start by considering a spin-orbital toy model containing   on-site relativistic spin-orbit coupling:
\bea
\label{toy}
H=\sum_j(\Delta \tau_j^z \tau_{j+1}^z-h \tau^z_j+ J \mathbf{S}_j \cdot \mathbf{S}_{j+1} -2 \lambda \tau^y_j S^z_j).
\eea
Here $\bm{\tau}$ is an orbital pseudospin-1/2 operator representing nearly degenerate $t_{2g}$ real orbitals ($|\tau^z=\frac12\rangle = |yz \rangle $ and $|\tau^z=-\frac12\rangle= | xz \rangle $\cite{chernperkins}) and $\mathbf{S}$ is a spin-1 operator.  The spins interact via exchange coupling $J$ and the orbitals via the Ising interaction $\Delta$, which reflects the directional nature of the orbitals. Physically, both $\Delta$ and $J$ originate from kinetic exchange processes in Mott insulators and are expected to be of the same order.  The orbital field $h$ depends on the strength of the structural distortion that lifts the orbital degeneracy. We consider $h$ as a parameter, controlled e.g. by pressure.
The last term  represents the projection of the spin-orbit interaction  $\lambda \mathbf{L} \cdot \mathbf{S}$, where $\mathbf{L}$ is the orbital angular momentum,   in the two-dimensional orbital subspace.

A ferro-orbital version of  model (\ref{toy}) was studied in Ref.~\onlinecite{Chern10}. Here we consider antiferro-orbital coupling $\Delta>0$, which arises naturally in the vanadates. \cite{tsunetsugu, chernperkins}  The antiferro-orbital model with  $h=0$ was studied in Ref. \onlinecite{nersesyan} using field theory methods. For $h=0$, it was found that in the regime of $J \gg \Delta$, a Gaussian transition separates a Haldane spin liquid AFO phase\cite{haldane}  at small $\lambda$ from a N\'eel spin-orbital phase at large $\lambda$. In the opposite regime of $ \Delta \gg J$, the Gaussian transition splits up into two Ising transitions where upon increasing $\lambda$ N\'eel spin order develops before the N\'eel order of $\tau^z$ orbitals disappears.
In the following we study the model at finite $h$, which is interesting because the orbital energy splitting competes with the Ising  interaction $\Delta$. We also focus on the limit of weak spin-orbit interaction, in which the excitations can be regarded as predominantly orbital or spin, as opposed to hybrid spin-orbital excitations. For $\lambda=0$, the spins are in the Haldane phase,which has an energy gap of about $\Delta_s\approx  0.41 J$,\cite{whitehuse}  and the orbitals are described as a classical  Ising chain.  The orbital sector exhibits a classical phase transition at $|h| =  \Delta$: for $|h|<\Delta$ the ground state has N\'eel order of $\tau^z$ (AFOz state) and for $|h|>\Delta$ it is a PO state  with orbitals polarized in the direction of the field (hereafter we assume $h \ge 0$).
The crucial aspect of this   transition is its macroscopic degeneracy: at the point $h=\Delta$, $\lambda=0$, all states that do not contain nearest-neighbor (nn) down pseudospins are degenerate. The number of configurations satisfying this constraint scales  like $\varphi^L$, where $\varphi=(1+\sqrt5)/2$ is the golden ratio and $L$ is the system size.~\cite{BaxterBook}

The macroscopic degeneracy of the classical critical point should be lifted by arbitrarily weak quantum fluctuations.
 In the regime $\lambda\ll \Delta,J$ the spin chain remains in the Haldane phase with a finite spin gap. On the other hand, for $h\approx \Delta$ the orbital gap becomes   small.  At energy scales far below the spin gap  the spin chain is in a singlet ground state
  and  the effective orbital model can be obtained by integrating out the spin fluctuations within second-order perturbation theory.

The procedure of integrating out the spin modes in the limit $\Delta\ll J$ was performed in Ref. \onlinecite{nersesyan} in the absence of the orbital field.
Here we do the same for finite $h$. We are mainly interested in the vicinity to the critical point, i.e. $\Delta \sim J \gg |\Delta - h|,\lambda$. Second order perturbation theory in the spin-orbit coupling in Eq. (\ref{toy}) yields an effective Ising orbital interaction~\cite{nersesyan}\be
\delta H = \sum_{j} \sum_{\ell \ge 1} (-1)^{\ell+1} J'_\ell \tau^y_{j} \tau^y_{j+\ell},\ee
where $J'_\ell \sim \frac{\lambda^2}{\Delta_s} e^{-\ell/\xi}$ is proportional to the spin-spin correlation in the Haldane ground state and decays exponentially for distances larger than the  correlation length $\xi \simeq 6$. For a qualitative understanding of the resulting orbital phases it suffices to consider the nn term $\ell=1$ as the leading transverse orbital coupling.
 The effective orbital model then becomes
\bea\label{ANNNI}
H_{\textrm{eff},\ell}=\sum_j(\Delta \tau_j^z \tau_{j+1}^z-h \tau^z_j+\kappa \tau_j^y \tau_{j+\ell}^y),
\eea
with $\ell=1$ and $\kappa\sim \lambda^2/\Delta_s \ll \Delta$. We keep the integer index $\ell$ in Eq. (\ref{ANNNI}) in order to refer to a family of models which will be convenient later. We recognize $H_{\textrm{eff},1}$ as the axial next-nearest-neighbor Ising (ANNNI) model in $1+1$ dimensions.  It has been extensively studied in statistical mechanics and finds applications, for instance, in  the physics of submonolayer adsorbates. For a review see, e.g., Refs. \onlinecite{selke,Chakrabarti}.
Based on various analytical and numerical methods it is known that the phase diagram  of the ANNNI model contains a multicritical point (MCP) at $h=\Delta$, $\kappa=0$, from which four phases emerge as shown in Fig. 1a.

\begin{figure}[b]
\begin{center}
\includegraphics*[width=85mm]{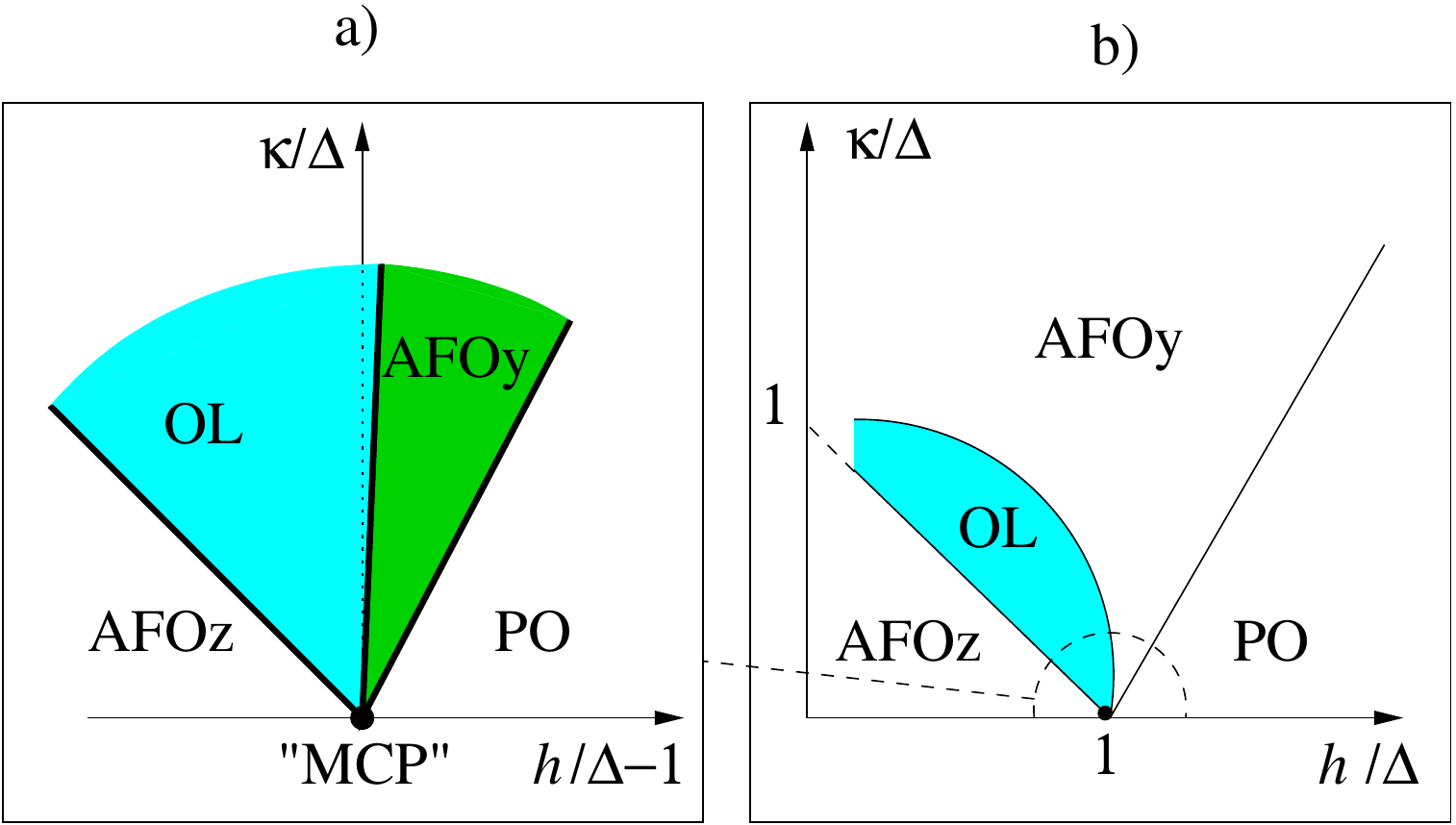}
\caption{(a) Phase diagram of effective orbital model~(\ref{toy}) near the
multicritical point  at $h=\Delta$, $\lambda=0$. Here $\kappa \propto \lambda^2 / \Delta_s$. The properties of the four orbital phases are described in Table \ref{table1}. (b) Schematic global phase diagram of the ANNNI model given in Eq.~(\ref{ANNNI}) with $\ell=1$.
 \label{fg:annni}}
\end{center}
\end{figure}

\begin{table}
\begin{center}
\begin{tabular}{|@{$\quad$}r@{$\quad$}r@{$\quad$}r@{$\quad$}r@{$\quad$} r@{$\quad$} |}
\hline
$~$ & $~$ & $\eta^z$ & $\eta^y$ &  degeneracy \\
\hline
$\uparrow \downarrow \uparrow \downarrow$ &AFOz &  $\ne 0$  & 0 &  2 \\
$\bullet \bullet \bullet \bullet$& OL & 0 & 0 &  1 (gapless)  \\
$\leftarrow \rightarrow \leftarrow \rightarrow$ & AFOy & 0 & $\ne 0$   &2 \\
$\uparrow \uparrow \uparrow \uparrow$ & PO & 0 & 0  & 1 \\
\hline
\end{tabular}
\end{center}
\caption{Symmetry characterization of the orbital phases occurring in the ANNNI model Eq.~(\ref{ANNNI}) with $\ell=1$. The phases are characterized in terms of the order parameters in Eq.~(\ref{eq:OP1}). For an intuitive picture we include cartoons of the orbital configurations, with $|\tau^z=\frac{1}{2} \rangle \equiv | \uparrow \rangle$, $|\tau^z=-\frac{1}{2} \rangle \equiv | \downarrow \rangle$, $|\tau^y=\frac{1}{2} \rangle \equiv | \rightarrow \rangle$,  $|\tau^y=-\frac{1}{2} \rangle \equiv | \leftarrow \rangle$, and $| \bullet \rangle$ denotes a pseudospin fluctuating in the $xy$ plane.} \label{table1}
\end{table}

The  phases are specified by broken symmetries with respect to  Eq.~(\ref{ANNNI}), namely: ($i$) translation and ($ii$) Ising Z$_2$ symmetry $\tau^y \to - \tau^y$, associated with order parameters
\bea
\label{eq:OP1}
\eta^z \equiv L^{-1}\sum_j(-1)^j\langle\tau^z_j\rangle,~~~\eta^y\equiv L^{-1}\sum_j(-1)^j\langle\tau^y_j\rangle,
\eea
 respectively.  The PO phase at large $h$ has a unique ground state and is disordered, $\eta^y=\eta^z=0$, whereas the AFOz phase is doubly degenerate and is characterized by finite $M_z$; see Table~\ref{table1} which also illustrates the pseudospin configuration. Two other phases appear between AFOz and PO. The   phase denoted AFOy, which is favored by large $\kappa$, breaks the Z$_2$ symmetry and has AFO order along $\hat{y}$,  hence finite $\eta^y$ (we assume $\kappa>0$). Generically, there cannot be a direct AFOy-AFOz transition within a Ginzburg-Landau theory of two independent order parameters, unless extra symmetry is present. Indeed, as implied by this symmetry argument, a so called ``floating phase'', here interpreted as an orbital liquid (OL), is known~\cite{Villain, Rujan,Allen} to arise in between the two; see Fig. 1. In this gapless phase pseudospins fluctuate in the $xy$ plane and have an  emergent U(1) symmetry corresponding to a conserved $z$-pseudospin magnetization.

Note, however, that if we extrapolate model (\ref{toy}) beyond the perturbative regime, an explicit  U(1) symmetry   occurs along the line $h=0$ at  $\kappa=\Delta$. At this point there is  a direct transition from the AFOz to the AFOy phase. The global phase diagram of the ANNNI model is shown schematically in Fig. 1b.~\cite{Villain, Rujan,Allen}  We believe that the U(1) symmetric point of the ANNNI model  is in correspondence with   the Gaussian transition discussed in Ref.~\onlinecite{nersesyan} for $h=0$ and $\Delta\ll J$, between the Haldane spin liquid AFOz phase at small $\lambda$ and the N\'eel spin-orbital phase at large $\lambda$. Indeed, in our AFOy phase finite $\eta^y$ orbital order implies a finite N\'eel order for the spins, $\langle \sum_i (-1)^i S^z_i \rangle \ne 0$, and the sign of this N\'eel order parameter is tied to the sign of $\eta^y$ such as to minimize the $\lambda$ term in Eq.~(\ref{toy}), hence the AFOy phase corresponds to the N\'eel spin-orbital phase.
An effective field theory was devised in order to access the various phases of the ANNNI model from the U(1) symmetric point $\kappa/\Delta-1=h/\Delta=0$.~\cite{Allen}  However, it remains unclear both from field theory~\cite{Allen} and from numerical studies~\cite{feo} whether or not the orbital liquid phase occurs already for infinitesimal $h$ from the Gaussian transition at $\kappa/\Delta=1$. Its existence was however confirmed numerically for $h/\Delta >0.05$.~\cite{feo} On the other hand the region of the phase diagram near the MCP $h=\Delta$, $\kappa=0$, has a transparent description in terms of effective hardcore particles,~\cite{Villain, Rujan} as will be discussed in the following subsection.

\subsection{Analysis of the ANNNI model close to the MCP\label{sec:analysis}}
In this section we briefly review field theory arguments that lead to the phase diagram of the ANNNI model in Fig. 1. Similar arguments will be used in the construction of the phase diagram for the zigzag model in Sec. \ref{se:Zigzag}.

For the ANNNI model, it is possible to obtain the transition lines near the MCP exactly. Following Ruj\'an,\cite{Rujan} we consider a deformation of the model (with $\ell=1$),
 \bea\label{XXZ1}
H_{\textrm{eff},\ell} &=&H_{\textrm{XXZ},\ell}+\delta H_\ell\nonumber\\
H_{\textrm{XXZ},\ell}&=&\sum_j \left[\frac{\kappa}{2} (\tau_j^x \tau_{j+\ell}^x+\tau_j^y \tau_{j+\ell}^y)+\Delta \tau_j^z \tau_{j+1}^z -h \tau_j^z \right],\nonumber\\
\delta H_\ell&=&\epsilon \sum_j \frac{\kappa}{2} (-\tau_j^x \tau_{j+\ell}^x+\tau_j^y \tau_{j+\ell}^y).
\eea
The ANNNI model is obtained by setting $\epsilon=1$. The model with  $\epsilon=0$ is the XXZ model, which  has a U(1) symmetry associated  with the conservation of  $\tau^z_{\textrm{tot}}=\sum_j \tau_j^z$. The XXZ model is exactly solvable by Bethe ansatz, but we shall follow a much simpler approach valid close to the MCP.\cite{Santos} We will use the XXZ model as starting point and  then analyze the effects of the $\epsilon$ perturbation in Eq. (\ref{XXZ1}).

It is convenient to introduce   Jordan-Wigner (JW) fermions by $\tau^z_j = \frac{1}{2}-\hat{n}_j$, $\hat{n}_j=c^\dagger_j c^{\phantom\dagger}_j$, $\tau^+_j=c^{\phantom\dagger}_j (-1)^j  e^{i \pi \sum_{l=1}^{j-1} \hat{n}_l}$. Then the XXZ model becomes\be H_{\textrm{XXZ},1}= \sum_j\left[ \Delta \hat{n}_j \hat{n}_{j+1} -\mu \hat{n}_j-\frac{\kappa}{4} (c_j^\dagger c^{\phantom\dagger}_{j+1}    + {\rm{h.c.}}) \right],\ee
and the perturbation reads \be \delta H_1 =\frac{\epsilon \kappa}{4} \sum_j c_j^\dagger c_{j+1}^\dagger + {\rm{h.c.}}.\ee
The chemical potential $\mu=\Delta-h$
 vanishes at the MCP. Near this point the largest energy scale is $\Delta$, which represents a strong repulsion between fermions occupying nn sites. Interestingly, in the limit $\Delta/\kappa\to \infty$, the XXZ model can be solved exactly \emph{without Bethe ansatz} by mapping to free fermions with an exclusion constraint.\cite{Santos}  The interaction is replaced by the hard-core constraint that no two fermions can occupy nn sites. Exactly at the MCP, all states which satisfy the nn constraint are degenerate and form a reduced  low energy subspace. The pairing operator $\delta H_1$ vanishes within  this subspace. To first order in $\kappa/\Delta$, the degeneracy is lifted by hopping of new fermions with annihilation operator $a_j$ on a reduced lattice with length $L_f= L(1-\nu)$ where $\nu=k_F/\pi$ is the filling fraction, described by the effective Hamiltonian~\cite{Santos}
 \bea
 H_{XXZ,1} = \sum_{j=1}^{L_f} \left[ -\frac{\kappa}{4}\left(a^\dagger_j a_{j+1}+a^\dagger_{j+1} a_{j} \right) - \mu a^\dagger_j a_{j} \right].
 \eea
 The ground state energy for fixed $\nu$ is easily evaluated,
 \bea
 \label{egs0}
 e_{gs}(\nu) &=&\lim_{L\to \infty}\frac{E_{gs}}{L} \nonumber \\
 &=& - \frac{\kappa}{2 L} \sum_{n=-\nu L/2}^{\nu L /2} \cos\left( \frac{2 \pi n}{L_f} \right)- \mu \nu .
 \eea
 Minimizing $e_{gs}(\nu) $ with respect to $\nu$, one finds~\cite{Trippe}
 \be
 \label{trippe}
 \frac{h}{\Delta}-1=\frac{\kappa}{2 \Delta} \left[ \frac{\cos \frac{\pi \nu}{1-\nu}}{1-\nu} -\frac{1}{\pi} \sin \frac{\pi \nu}{1-\nu} \right].\ee
 As a result, the MCP of the XXZ model (in the absence of $\delta H_1$) is a triple point with  a critical region in which $\nu$ changes continuously in the phase diagram for $-\frac{\kappa}{\Delta} < \frac{h}{\Delta}-1 < \frac{\kappa}{2 \Delta}$,
separating the AFOz with $\nu=1/2$ from the PO phase with $\nu=0$. The low energy physics of this orbital liquid (OL) phase is described by the Luttinger model\cite{Giamarchi}
\bea\label{HLL}
H_{LL}=\frac{v}{2} \int dx  \left[K(\partial_x \theta)^2+K^{-1}(\partial_x \phi)^2 \right],
\eea
with $[\phi(x),\partial_{x'} \theta(x')]=i\delta(x-x')$, and the bosonization formula of JW fermions is $c_j \to c(x)\sim e^{i k_F x} e^{- i \sqrt{\pi} (\theta-\phi)} +e^{-i k_F x} e^{- i \sqrt{\pi} (\theta+\phi)} $. Here $v$ is the velocity, and $K$ is the Luttinger  parameter. The velocity $v$ is determined by the energy spacing for a finite size system $\Delta_L=\frac{2\pi}{L}v$. From Eq.~(\ref{egs0}) we find $\Delta_L=\frac{\pi \kappa}{L (1-\nu)} \sin \left( \frac{\pi \nu}{1-\nu} \right)$ giving $v=\frac{\kappa}{2(1-\nu)} \sin \left( \frac{\pi \nu}{1-\nu} \right)$. The Luttinger parameter is related to the compressibility by~\cite{Giamarchi} $K=\pi v \left(\frac{\partial^2 e_{gs}}{\partial \nu^2} \right)^{-1}$. Using Eq.~(\ref{egs0}), one finds to first order in $\kappa/\Delta$ that $K=(1-\nu)^2$.~\cite{Santos} The transition to the PO state in the XXZ model corresponds to the limit of vanishing density, $\nu \to 0$, where $K\to 1$. The transition to the AFOz state corresponds to half-filling $\nu \to 1/2$, in which case $K\to 1/4$. The latter is a commensurate-incommensurate (C-IC) transition.\cite{Giamarchi}

The phase diagram of the ANNNI model can be interpreted in terms of the instability of the critical phase against  $\delta H_1$ (for general $0 < \epsilon \le 1$). Within  second order perturbation theory,  besides fermion-fermion interactions, a next-nearest-neighbor (nnn) pairing term $\delta H_{\textrm{nnn}}\sim (\kappa^2/\Delta) \sum_j (c_j c_{j+2}+{\rm{h.c.}})$ is generated in the reduced low energy subspace.
 In the bosonic language, the leading U(1) symmetry  breaking perturbation is the operator $\cos (2 \sqrt{\pi} \theta)$.\cite{Allen} In our notation the latter has scaling dimension $1/K$ and becomes relevant when $K>1/2$.  Since $K=1/4$ at the  C-IC transition, $\delta H_{1}$ is irrelevant near the boundary with the AFOz phase and the system has an emergent U(1) symmetry in the OL phase. Moving towards larger $h$,  a Kosterlitz-Thouless transition takes place at the line where $K=1/2$,
 which can be  found using Eq.~(\ref{trippe}) at $\nu=1-1/\sqrt{2}$.  Beyond this line the U(1) symmetry breaking term $\delta H_1$ opens up a gap in the spectrum. The $\theta$ field is pinned, corresponding to long-range order in the plane perpendicular to the $z$ direction. For $\kappa>0$ in Eq. (\ref{ANNNI}), this implies N\'eel order of $\bm{\tau}_j$ along the $y$ direction, with order parameter  $\eta^y \ne 0$.
 Finally, if we keep increasing $h$, an Ising transition into the polarized PO state takes place.

 We note that adding further neighbor interactions to Eq.  (\ref{ANNNI}), one can argue that as long as these can be treated as small perturbations, the phase diagram in Fig. \ref{fg:annni} remains qualitatively correct. The perturbations renormalize the parameters of the effective field theory and the transition lines can no longer be determined exactly, but the Luttinger parameter still assumes the universal value $K=1/4$~\cite{Giamarchi} at the transition to the AFOz phase, and the OL is stable against U(1) symmetry breaking perturbations. However, when  further neighbor  interactions dominate, the phase diagram can change qualitatively. In any case, it is guaranteed by the macroscopic degeneracy of the critical point that new phases will emerge above it. The case of dominant nnn interactions will be discussed in  Sec. \ref{se:Zigzag}.

\section{Zigzag  model}
\label{se:Zigzag}
We now carry out the same program  as in Sec. \ref{se:toy} for the spin-orbital model for CaV$_2$O$_4$\cite{chernperkins}
\bea\label{Hzz}
H_{ZZ} &=& \sum_j [\Delta \tau_j^z \tau_{j+1}^z -h \tau_j^z-2 \lambda S_j^z \tau_j^y+  J_2 \mathbf{S}_j \cdot \mathbf{S}_{j+2} \nonumber \\
&&
+ (J_1 \bar{\mathcal{O}}_j  -J_1' \mathcal{O}_j ) \mathbf{S}_j \cdot \mathbf{S}_{j+1} ],
\eea
where $J_1,J_1',J_2,\Delta>0$. Compared to the toy model in Eq.~(\ref{toy}), here the nn spin exchange interaction depends on the orbital configuration (Kugel-Khomskii type interactions)
via the antiferro and ferro-orbital bond operators
$
\mathcal{O}_j=\frac{1}{2}(1-4 \tau^z_j \tau^z_{j+1})$ and $
\bar{\mathcal{O}}_j=\left[ \frac{1}{2}+(-1)^j \tau_j^z \right] \left[ \frac{1}{2}+(-1)^j \tau_{j+1}^z\right]
$.
In the zigzag structure the nnn interaction $J_2$ arises due to direct exchange of occupied $|xy \rangle$ orbitals.
Experiments~\cite{pieper} on CaV$_2$O$_4$
suggest  that $J_2 \gg J_1 \gg J_1'$, implying that spins are close to the limit of weak coupling between even and odd sublattices.

We start with symmetry considerations. Model (\ref{Hzz}) has the following discrete symmetries: ({\it i}) time reversal symmetry $\mathbf{S} \to - \mathbf{S}$, $\tau^{x,z}\to\tau^{x,z}$, $\tau^y\to -\tau^y$; ({\it ii}) reflection about either even or odd bond centers: $j\to -j+1$ or $j \to -j -1$.
Note that for $h>0$  the nonzero orbital magnetization $L^{-1}\langle\sum_j\tau_j^z\rangle>0$, together with the staggered term in the ferro-orbital bond operator  $\bar{\mathcal{O}}_j$,  imply that nn even bonds are different than  odd bonds. As a result,  the model has translational invariance only by two sites. Only at $h=0$ is the Hamiltonian  invariant under the transformation that combines translation by one site and orbital inversion $\tau^z \to - \tau^z$.

For $J_2\gg J_1,J_1^\prime, \lambda$, we expect the spin chain to have a finite Haldane gap, close to the value for decoupled Haldane chains $\Delta_s \approx 0.41 J_2$. Assuming that this is the case, we take the expectation value of spin operators in the ground state in order to write down the effective orbital model. The spin correlations are strongest between nnn. Between nn, the correlation is weaker and oscillates due to  the field-induced orbital dimerization, $\langle  \mathbf{S}_j \cdot \mathbf{S}_{j+1} \rangle =C [1+(-1)^j A]$, with $C\sim J_1/J_2 \ll 1$. The   transverse orbital couplings are generated at second order in the spin-orbit coupling. The low energy effective orbital model becomes
\be
\label{effective}
H_{ZZ}^{{\textrm{eff}}}=\sum_j \{\tilde{\Delta} [1+\zeta (-1)^j] \tau^z_j \tau^z_{j+1}-\tilde{h} \tau^z_j +\kappa_2 \tau^y_j \tau^y_{j+2}\},
\ee
where $\tilde{\Delta}=\Delta+C(J_1+2 J_1')$, $\zeta =C A (J_1+2 J_1')/\tilde{\Delta} \ll 1$, $\tilde{h}=h-C A J_1$, and $\kappa_{2}\sim \lambda^2/\Delta_s$.
In Eq. (\ref{effective}) we have neglected orbital interactions beyond nnn as well as terms of higher order in the inverse spin gap. Among these is the nn transverse coupling, $\kappa_1\tau^ y_j\tau^y_{j+1}$, with $\kappa_1\sim C \lambda^2/\Delta_s$.

For $ \kappa_2=0$, Eq. (\ref{effective}) reduces to a classical Ising chain with a critical point at $\tilde{\Delta}=\tilde{h}$ separating the AFOz from the PO state as long as $|\zeta|<1$. For any  value of  $|\zeta|<1$, at the critical point all classical states satisfying the constraint of   no nn down pseudospins are degenerate. Setting $\zeta=0$, we obtain $H_{\textrm{eff},2}$ in Eq. (\ref{ANNNI})  with coupling constants $(\Delta,h,\kappa)\to(\tilde{\Delta},\tilde{h},\kappa_2)$. Note that by setting $\kappa_1=0$ in model (\ref{effective}) we have artificially  ({\it i}) restored the  translation symmetry by one site and ({\it ii})  enlarged the Ising symmetry from Z$_2$ to Z$_2\otimes$Z$_2$, leading to two Ising order parameters
\be
\label{eq:OP2}
\eta^y_m \equiv L^{-1}\sum_j(-1)^j\langle\tau^y_{2j+m}\rangle,
\ee
where $m=0,1$ is the index for even and odd sublattices, respectively. However this enlarged symmetry will not change the  phase diagram in the region of interest near the MCP, namely for $\kappa_2\ll \tilde{\Delta}\approx \tilde{h}$.

\section{Exact solution of a U(1)$\otimes$U(1) symmetric XXZ model}
\label{se:Exact}

To our knowledge, the phase diagram of  $H_{\textrm{eff},2}$ has not been studied. In order to examine the possibility of a critical phase we again consider the deformation of the model in Eq.~(\ref{XXZ1}), this time with $\ell=2$, by perturbing around a U(1)$\otimes$U(1) symmetric   model. The point of interest is $\epsilon=1$, but we start from $\epsilon\ll 1$.

\begin{figure}[b]
\begin{center}
\includegraphics*[width=65mm]{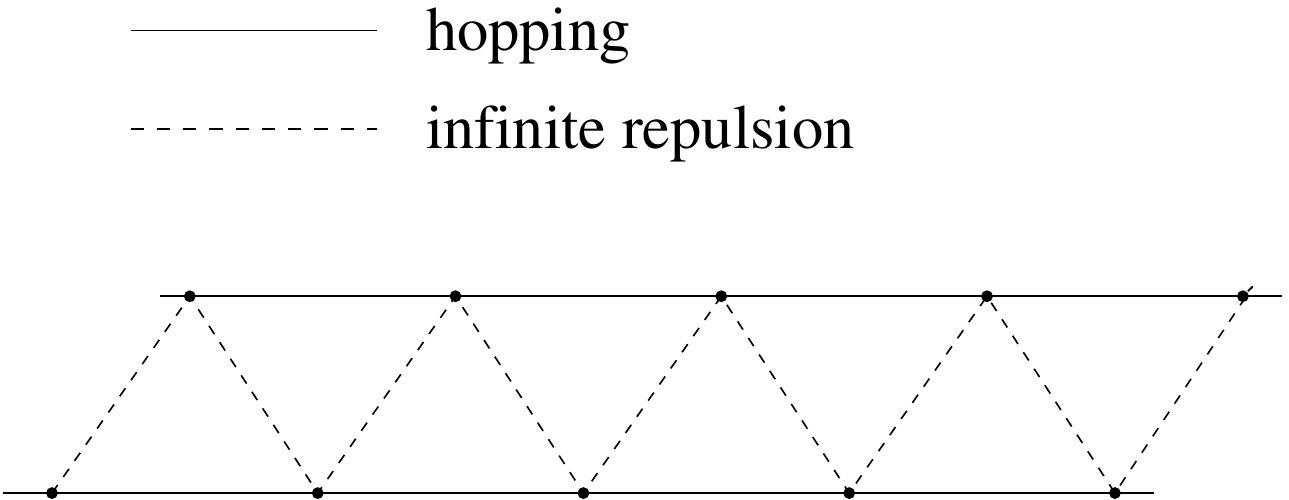}
\caption{Schematic representation of the exactly solvable U(1)$\otimes$U(1) symmetric zigzag model with    $\Delta/\kappa\to\infty$, corresponding to spinless fermions hopping on either even or odd sublattices, with infinite repulsion along zigzag bonds.
 \label{fg:zz}}
\end{center}
\end{figure}

In terms of JW fermions, the model with  $\epsilon=0$ near the MCP describes fermions hopping along the legs of a zigzag ladder,
\begin{equation}
H_{{\rm{XXZ}},2}=\sum_{j=1}^{L} \left[\frac{\kappa}{4}(c^\dagger_jc^{\phantom\dagger}_{j+2}+ \textrm{H.c.}) - (\Delta-h) c^\dagger_j c_j\right].
\end{equation}
For $\Delta/\kappa\to\infty$, the strong nn repulsion can be replaced by  the local  constraint $ \hat{n}_j \hat{n}_{j+1} =0$ for all $j$; see Fig.~\ref{fg:zz}.

\begin{figure}[b]
\begin{center}
\includegraphics*[width=65mm]{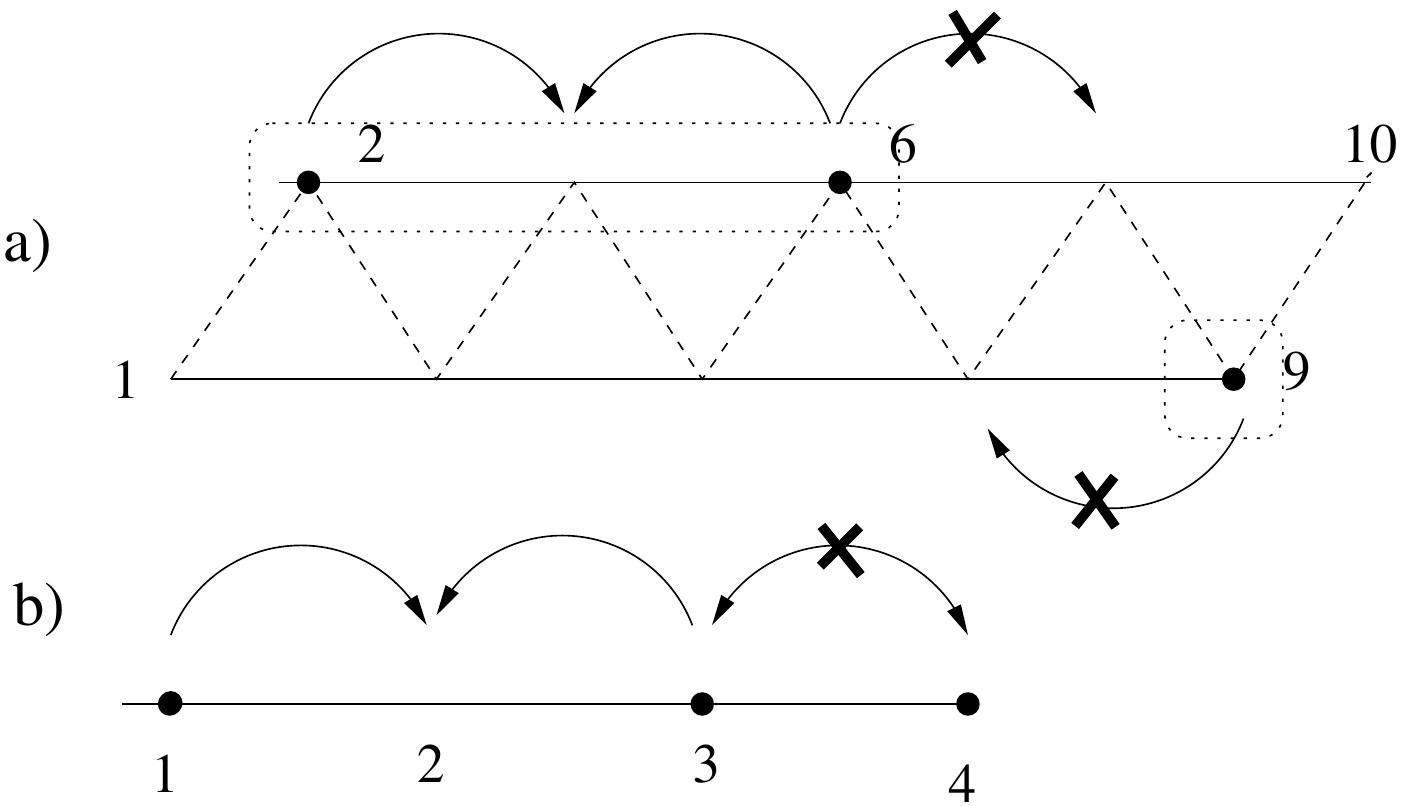}
\caption{a) Example of a configuration with $N_d=2$ domains in the original zigzag chain, with $n_1=2$ fermions in the first domain at positions $r_{1,1}=2$, $r_{1,2}=6$, and with $n_2=1$ fermion in the second domain located at position $r_{2,1}=9$. b) At the fictitious lattice with length $L'=4$, constructed by keeping only the  sites inside   dashed rectangles in (a), the occupied sites are $r'_{1}=1$, $r'_{2}=3$, and $r'_{3}=4$. In the original lattice there are only two possible hopping processes denoted by arcs without crosses, compatible with the infinite repulsion along zigzag bonds.  The interaction forbidden hopping process are blocked in the fictitious lattice due to the Pauli principle.
 \label{fic}}
\end{center}
\end{figure}

Due to the constraint and the absence of nn hopping ($\kappa_1=0$), any fermion configuration  can be classified according to alternating domains of fermions hopping on even and odd sublattices: starting from the left end of an open chain, we find first $n_1$ fermions on the even sublattice, then $n_2$ fermions on the odd sublattice, then $n_3$ fermions on the even sublattice, and so on (for definiteness we assume $n_1\ne 0$ for  states with nonzero fermion filling $\nu=N/L$, and $L$ even). Importantly, besides the trivial conservation of the number of fermions on even and odd sublattices, we have a set of good quantum numbers $n_\alpha $, $\alpha=1,2,...N_d$, where $N_d$ is the number of domains for open boundary conditions.
We now define fictitious fermions with annihilation  operator $a_{r'}$ that act on the Hilbert space of a fictitious   lattice of reduced length.  Let $r_{\alpha,i}$ denote the position of the \protect{$i$-\textit{th}} ($1 \le i \le n_\alpha$) fermion in the \protect{$\alpha$-\textit{th}} domain in the original zigzag chain, measured from the left.  This corresponds to the \protect{$i^\prime$-\textit{th}} fermion, $i^\prime=\sum_{\alpha^\prime<\alpha}n_{\alpha^\prime}+i$, in the zigzag chain. The reduced lattice is defined such that the position of the  \protect{$i$-\textit{th}}  fermion in the \protect{$\alpha$-\textit{th}} domain in the original lattice gets mapped to\be
r_{\alpha,i}\to r^\prime_{i^\prime}=(r_{\alpha,i}-\alpha+1)/2.
\ee
The length of the fictitious lattice is   $L^\prime=(L-N_d+1)/2$ for $N_d$ odd or $L^\prime=(L-N_d)/2$ for $N_d$ even. For an example of a configuration  on the original and fictitious lattices, see Fig.~\ref{fic}.

It is the Pauli exclusion for the $a_{r^\prime}$ fermions that selects, after the prescribed mapping, only the states which satisfy the nn constraint in the original system. Accordingly, we find that the effective Hamiltonian in the low energy subspace is noninteracting: \bea
H_{{\rm{XXZ}},2}&=&\sum_{j=1}^{L^\prime} \left[\frac{\kappa}{4}(a^\dagger_ja^{\phantom\dagger}_{j+1}+\textrm{H.c.})- (\Delta-h) a^\dagger_j a_j \right],\nonumber \\
\Delta/\kappa &\to& \infty.
\eea
Therefore, the ground state energy can be promptly evaluated as a function of fermion density $\nu$ and domain wall density $q\equiv(N_d-1)/L$; we find
\bea
e_{gs}(\nu,q)=
(h-\Delta)\nu-\frac{\kappa}{4\pi}(1-q)\sin\frac{2\pi\nu}{1-q}.\label{egs}
\eea
Minimizing with respect to $\nu$ at fixed $h$, we find that for $|h- \Delta|<\kappa/2$ the ground state has a finite density of fermions
\be
\nu_0=\frac1{2\pi}\arccos\frac{ h- \Delta}{\kappa/2},
\ee
but zero density of domain walls. The chemical potential for domain walls \be \mu_d=(\partial e_{gs}/\partial q)|_{\nu_0}=(\kappa/4\pi)\sin 2\pi\nu_0-( h-\Delta)\nu_0\ee
vanishes only for $\nu_0= 0$.  Thus the $c$ fermions prefer to occupy a single sublattice, giving a finite value of $\eta^z$.  Therefore, the ground state of $H_{{\rm{XXZ}},2}$ in the limit $\Delta/\kappa\to \infty$ for $| h- \Delta|<\kappa/2$ is a LL  that breaks    bond reflection symmetry. Only intrachain correlations within the partially occupied sublattice decay algebraically.  Notice that the nn $\kappa_1$ process mentioned below Eq. (\ref{effective}) involves interleg hopping, which creates domain walls. Due to the finite energy cost for domain walls, the liquid phase is stable against small  $\kappa_1$ perturbations.

\begin{figure}[b]
\begin{center}
\includegraphics*[width=85mm]{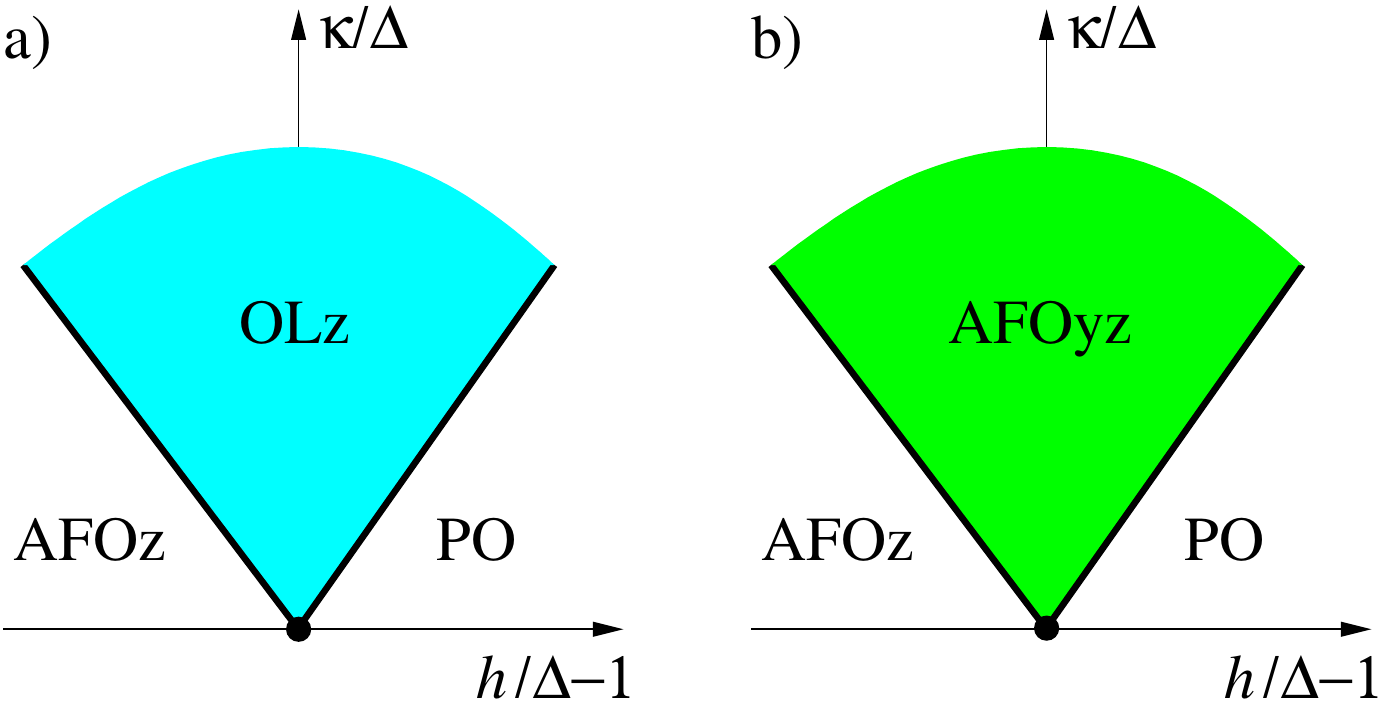}
\caption{a) Phase diagram of the   U(1)$\otimes$U(1) symmetric  model $H_{\textrm{XXZ},2}$ in Eq. (\ref{XXZ1}). The symmetry properties of the phases are given in table~\ref{table2}. b) Adding an infinitesimal $\epsilon$ perturbation in $H_{{\rm{eff}},2}$ in Eq.~(\ref{XXZ1}) leads to the opening of an  energy gap in the OLz phase of a); the resulting phase AFOyz has coexisting order parameters as specified in table~\ref{table2}.
 \label{fg:OLz}}
\end{center}
\end{figure}

The phase diagram of $H_{\textrm{XXZ},2}$  near the MCP is plotted in  Fig. \ref{fg:OLz}a, and the symmetry properties of the three phases that appear around the MCP are given in Table~\ref{table2}. The LL phase with finite $\eta^z$ occurring for $|h- \Delta|<\kappa/2$ is denoted OLz. Its cartoon picture corresponds to pseudospins polarized along the field $(\uparrow)$ in one sublattice and fluctuating in the $xy$ plane for the other sublattice; see Table \ref{table2}. For $h- \Delta>\kappa/2$, Eq.~(\ref{egs}) is minimized at $\nu_0=0$, giving the PO phase with its exact classical ground state shown in Table~\ref{table2}. For $h- \Delta<-\kappa/2$, Eq.~(\ref{egs}) is minimized at $\nu_0=1/2$, giving the AFOz phase.

\begin{table}
\begin{center}
\begin{tabular}{|@{$\quad$}r@{$\quad$}r@{$\quad$}r@{$\quad$}r@{$\quad$} r@{$\quad$} r@{$\quad$}|}
\hline
$~$ & $~$ & $\eta^z$ & $\eta_0^y$ & $\eta_1^y$ & degeneracy \\
\hline
$\uparrow \uparrow \uparrow \uparrow$ & PO & 0 & 0 & 0 & 1 \\
$\uparrow \rightarrow \uparrow \leftarrow$ & AFOyz & $\ne 0$ & $\ne 0$(0) & 0($\ne 0$) &4 \\
$\uparrow \downarrow \uparrow \downarrow$ &AFOz &  $\ne 0$ & 0 & 0 & 2 \\
$\uparrow \bullet \uparrow \bullet$& OLz & $\ne 0$ & 0 & 0 & 2 (gapless) \\
$\rightarrow \rightarrow \leftarrow \leftarrow$ &AFOy4 & 0 & $\ne 0$ & $\ne 0$ & 4\\
\hline
\end{tabular}
\end{center}
\caption{Symmetry characterization of the phases stabilized by the effective orbital model in Eq. (\ref{XXZ1}) with $\ell =2$. For an intuitive picture we include cartoons of the orbital configurations, with $|\tau^z=\frac{1}{2} \rangle \equiv | \uparrow \rangle$, $|\tau^z=-\frac{1}{2} \rangle \equiv | \downarrow \rangle$, $|\tau^y=\frac{1}{2} \rangle \equiv | \rightarrow \rangle$,  $|\tau^y=-\frac{1}{2} \rangle \equiv | \leftarrow \rangle$, and $| \bullet \rangle$ denotes a fluctuating orbital. The phases are characterized in terms of the order parameters in Eqs.~(\ref{eq:OP1}) and (\ref{eq:OP2}).} \label{table2}
\end{table}

Next, we consider the effect  of a weak perturbation  $\epsilon\ll1$ in Eq.~(\ref{XXZ1}) with respect to the LL fixed points in  the OLz phase. Since domain walls have a finite energy cost $\mu_d$, they are not created at small enough $\epsilon$. Thus $\langle \hat{n}_{2j+m} \rangle =0$  for one sublattice, $m=0$ or 1 (hence $\eta^z \ne 0$ still holds). Setting $\langle \tau^z_{2j+m} \rangle=1/2$ in Eq.~(\ref{XXZ1}), we obtain an exactly solvable  anisotropic XY model for the other  $\bar{m} = 1-m$ sublattice,
\bea
H_{\textrm{XY}}&=&\sum_j \left[(\Delta-h) \tau_{2j+\bar{m}}^z +\frac{\kappa(1-\epsilon)}{2} \tau_{2j+\bar{m}}^x \tau_{2j+2+\bar{m}}^x \right.\nonumber \\
&&+
\left.\frac{\kappa(1+\epsilon)}{2} \tau_{2j+\bar{m}}^y \tau_{2j+2+\bar{m}}^y \right], \label{XYmodel}
\eea
with $\epsilon \to 0^+$.
It is readily found that an energy gap opens for finite $\epsilon$ in the region previously occupied by the gapless OLz phase. Physically, the $\epsilon$ perturbation lowers the U(1)$\otimes$U(1) symmetry down to Z$_2\otimes$Z$_2$. The Ising Z$_2$ symmetry acting on the fluctuating sublattice $\bar{m}$ of the OLz phase is now broken, resulting in a phase with $\eta_{\bar{m}}^y \ne 0$. This result also follows   from the field theory analysis of the instability of the  LL   under the U(1) symmetry breaking perturbation discussed in Sec. \ref{sec:analysis}: it can be   verified by taking derivatives of the ground state energy that the Luttinger parameter of the OLz phase is $K=1$ (free fermions), independent of $\nu$. As a result, U(1) symmetry breaking is a relevant perturbation for all values of $\nu$. This is in contrast with the result for $H_{\textrm{XXZ},1}$;\cite{Santos} the crucial difference is that the size of the reduced lattice defined here for the zigzag chain does not depend on the number of fermions, only on the number of domains, which is always $N_d=1$ for the ground state.  The resulting phase diagram at fixed infinitesimal $\epsilon$ is shown in Fig. \ref{fg:OLz}b. The phase that replaces the OLz is denoted AFOyz due to the coexisting $\eta^z$ and $\eta_{\bar{m}}^y$ order parameters; see also Table \ref{table2}.

The effective   XY model (\ref{XYmodel}) applies also along the transition from AFOyz to the  AFOz phase. Physically, even at finite $\epsilon$ we expect the AFOz--AFOyz transition to be of Ising type and to occur within one sublattice, with the other sublattice being essentially frozen by the field. However, the transition to the PO phase, at which the gap for the creation of domain walls vanishes, $\mu_d \to 0$, needs further consideration at finite $\epsilon$. For this reason, we shall now expand our treatment of the model to the global phase diagram, based on general field theory arguments.

\section{Global phase diagram of Z$_2\otimes$Z$_2$ coupled Ising chains}
\label{se:Global}
\begin{figure}[b]
\begin{center}
\includegraphics*[width=55mm]{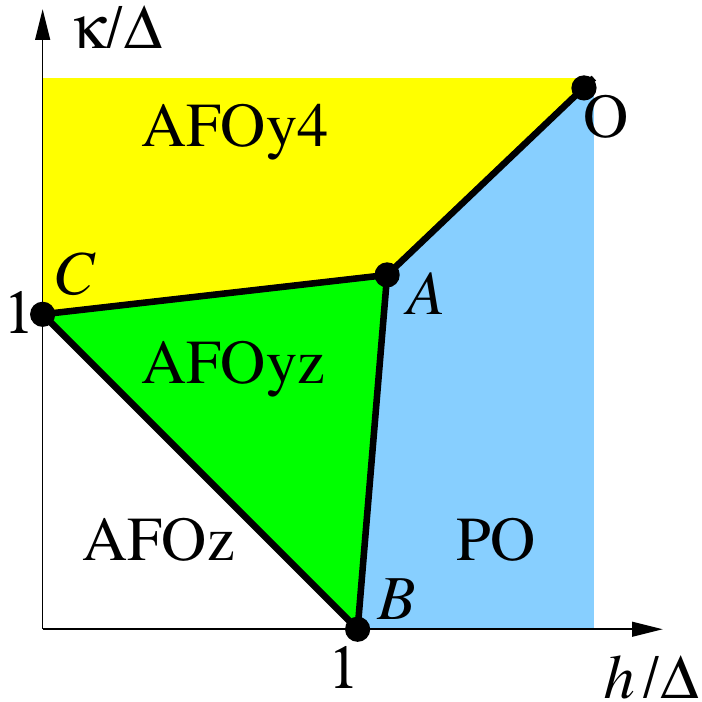}
\caption{Schematic global phase diagram of model (\ref{ANNNI}) with $\ell=2$, namely with NNN transverse couplings. All phases are gapped and characterized according to the broken symmetries as given in Table \ref{table2}. Starting from the disordered PO phase, both Ising order parameters $\eta_0^y$ and $\eta_1^y$ acquire finite values upon crossing line OA. In the AFOyz phase the $\eta^z$ order parameter is finite, as well as one of the two $\eta^y_m$ ($m=0,1$) order parameters. All three critical lines emanating from point A are described by a Luttinger liquid theory, which follows from a formal analogy between the effective theory of point A, Eq.~(\ref{eff}), and the effective theory of the XYZ spin chain Eq.~(\ref{eq:XYZ}). Point A has an emergent SU(2) symmetry, corresponding to the Heisenberg point of the XYZ spin chain, and realized in terms of the dual $\mu$ variables defined in Eq.~(\ref{eq:mu}).
 \label{fg:global}}
\end{center}
\end{figure}

In this section we argue that the global phase diagram of $H_{{\rm{eff}}, 2}$ in Eq.~(\ref{ANNNI}) has the form shown in Fig. \ref{fg:global}. One of our motivations is to understand whether there is a direct transition between the PO and AFOyz phases near the MCP.

The main argument is based on the effective  field theory starting in the limit $\kappa\sim h\gg \Delta$. For $\kappa/ h$ finite  and $h/ \Delta  \to \infty$,  $H_{{\rm{eff}}, 2}$ in Eq.~(\ref{ANNNI})  is equivalent to two decoupled transverse field Ising chains on the even and odd sublattices. At large $\kappa/h$,  we expect a fourfold degenerate phase, denoted AFOy4, which breaks both Z$_2$ Ising symmetries; see Table \ref{table2}.  If we decrease the ratio $\kappa/h$ keeping $h/\Delta$ large, there is a transition from this ordered phase   to the PO phase on a line given asymptotically by $h = \kappa/2$. Point  O in Fig. \ref{fg:global}  represents the limit $h = \kappa/2$, $\kappa/\Delta\to \infty$. At this point the AFOy4-PO transition is described by two decoupled Majorana fermion theories  with Hamiltonian density \cite{zuber}
\be\mathcal{H}_0=i \sum_{p=1}^2  \left[\frac{v_0}{2}  (\chi_{pL } \partial_x \chi_{pL}-\chi_{pR } \partial_x \chi_{pR})+  m_p  \chi_{pL} \chi_{pR}\right],\ee
with $p$ the sublattice index,  velocity $v_0 = \kappa/2 $ and masses $m_1=m_2=h -\kappa/2$.

However, the theory of the generic AFOy4--PO transition, $\mathcal{H} =\mathcal{H}_0 + \mathcal{H}_\Delta$, includes a correction  due to  the coupling $\propto \Delta$ between the two Ising chains.  Since in the continuum limit~\cite{zuber}
  \be
  \label{Mzcont}
  \tau^z_{2 x + p}\sim i \chi_{p L}(x) \chi_{pR}(x),
   \ee
   we obtain \be\mathcal{H}_\Delta =-2 \Delta    \chi_{1 L} \chi_{1 R}  \chi_{2 L} \chi_{2 R}.\ee Defining one Dirac fermion with left and right moving components, \be
   \label{Diracpsi}
   \psi_{L,R}=\frac{\chi_{1 L,R}+ i \chi_{2 L,R}}{\sqrt{2}},\ee
we may decompose the operator $\mathcal{H}_\Delta = \frac{\Delta}{2}(\psi_L + \psi_L^\dagger) (\psi_R + \psi_R^\dagger) (\psi_L - \psi_L^\dagger) (\psi_R - \psi_R^\dagger)$ into a list of operators. At small $\Delta / \kappa$ the most important contribution is the marginal attractive interaction $\sim - \psi^\dagger_L \psi_L \psi^\dagger_R \psi_R$. Consequently, the AFOy4-PO transition becomes a LL theory with continuously varying exponents, as encountered in the Ashkin-Teller model.\cite{AT}
Upon increasing $\Delta/\kappa$ further, the first operator  that becomes relevant is $\sim \psi_L \psi_L \psi_R \psi_R$. Bosonizing~\cite{Giamarchi} with  $\psi_{L,R}=\frac{1}{\sqrt{2 \pi}} e^{-i \sqrt{\pi} (\phi \pm \theta)} $, we obtain
\bea
\label{eff}
\mathcal{H}&=&  \frac{v}{2} [K(\partial_x \theta)^2+K^{-1} (\partial_x \phi)^2] \nonumber \\
&&+ g_0 \cos(\sqrt{4 \pi} \theta) + g_u \cos(4 \sqrt{ \pi} \phi),
\eea
with $g_0 \propto m_1=m_2$, $g_u = \frac{\Delta}{8 \pi^2}$, renormalized velocity $v=v_0+\mc{O}(\Delta^2)$, and $K \approx 1-\frac{\Delta}{\pi v_0}$.
 [We ignored irrelevant terms like $\psi_{L/R}^\dagger \psi_{L/R} \psi_{R/L} \psi_{R/L} \sim \partial_x (\phi \pm \theta) e^{-i\sqrt{4 \pi}(\phi \mp \theta)}$ and $\psi_L^\dagger \psi_L^\dagger \psi_R \psi_R \sim e^{i4 \sqrt{\pi} \theta}$.]
 Notice the bosonization formula used here differs from the one in Eq. (\ref{HLL}) by a duality transformation $\theta\leftrightarrow\phi$. The $g_0$ operator has scaling dimension $1/K$ and is relevant for $K>1/2$. In this representation, relevant $g_0>0$ describes the  massive AFOy4 phase and relevant $g_0<0$ the PO phase. The critical line approaching O is defined by setting $g_0=0$.

The $g_u$ operator in Eq. (\ref{eff}) has scaling dimension $4K$. As we depart from point O along the critical line, the LL parameter $K$ decreases below unity. At the point where $K$ reaches the value $1/2$, denoted by A in Fig.  \ref{fg:global},   the  $ g_u$  perturbation becomes relevant and the system undergoes a Kosterlitz-Thouless transition into a gapped phase. To identify this gapped phase, recall that both Z$_2$ Ising orders are on their instability along the OA line, $m_1 = m_2 =0$. Upon crossing point A the field $\phi$ gets locked in the minima of the $g_u$ cosine term, hence the order parameter [using Eqs.~(\ref{Mzcont}), (\ref{Diracpsi}) and $\psi_{L,R}=\frac{1}{\sqrt{2 \pi}} e^{-i \sqrt{\pi} (\phi \pm \theta)} $]
\be
\eta^z \sim  i \chi_{1L}  \chi_{1R} - i \chi_{2L}  \chi_{2R}    \sim \sin(\sqrt{4 \pi} \phi)
 \ee
 becomes finite. This implies opposite expectation values for the two mass terms in $\mathcal{H}_0$. Equivalently, within a simple  mean field picture valid at finite $\eta^z$, the interaction $\mathcal{H}_\Delta$ gives rise to  a finite mass  difference $m_1 -  m_2 \propto \eta^z$.
 As one mass term becomes positive and the other one negative, one Ising Z$_2$ symmetry is spontaneously broken, leading to $\eta^y_m \ne 0$ for $m=0$ or $1$. Therefore, upon crossing A along OA, we reencounter  the AFOyz phase discussed in Sec. \ref{se:Exact}.

Interestingly, the field theory in Eq.~(\ref{eff}) controlling point A coincides with the field theory of the XYZ model
\be
\label{eq:XYZ}
H_{XYZ}=\sum_j \sum_{a=x,y,z} J_a  S^a_j S_{j+1}^a
\ee
near the Heisenberg point $J_x=J_y=J_z$.\cite{Giamarchi} In the XYZ model  three gapped N\'eel phases with staggered magnetization along $a=\hat{x},\hat{y}$ or $\hat{z}$ meet at the Heisenberg point, and are separated by LL transitions occurring when the two largest $J_a$'s are equal (see also Ref.~\onlinecite{sela2011}). From this field theory equivalence, we conclude that the LL transition AO splits at point A into two LL transitions; see Fig. \ref{fg:global}.  It is natural to infer a LL type AFOyz--PO transition extending all the way to the MCP, connecting points A and B as in Fig. \ref{fg:global}.

Defining dual variables (normalized as Pauli matrices),
\be
\label{eq:mu}
2 \tau_j^z = -\mu_j^x \mu_{j+1}^x,~~2 \tau_j^y = \prod_{l \le j} \mu_l^y,
\ee
in terms of which the model becomes \be
H_{{\rm{eff}}, 2}=\frac{1}{4}\sum_j(\Delta \mu^x_j \mu_{j+2}^x+\kappa \mu^y_j \mu_{j+1}^y+2 h \mu^x_j \mu_{j+1}^x), \ee
we see that our model is self-dual along the line  $h=0$. As we expect a single transition along the $\kappa$ axis, separating AFOz and AFOy4, this point denoted $C$ in Fig. \ref{fg:global} must occur at $\kappa = \Delta$.  It is natural to connect this point with the third LL line coming out of $A$. Finally, we expect an Ising transition extending  from $C$ to $B$ and separating the AFOz and AFOyz phases, which differ by a single Ising order parameter.

We mention that bosonizing the dual model   using  $\mu^z_j \sim \frac{2}{\sqrt{\pi}} \partial_x \phi$
  we obtain back Eq.~(\ref{eff}). In terms of dual variables the  AO line is an XY model, and the perpendicular direction  corresponds to finite  $J_x-J_y$. At point A the $g_u$ perturbation, which  corresponds to the  umklapp operator in spin chain language,\cite{Giamarchi} becomes relevant. The value $K=1/2$ at point A implies an emergent SU(2) symmetry~\cite{Giamarchi}  manifested in the correlation function for the components of the $\bm{\mu}$ vector.

The global phase diagram changes upon including small finite nn transverse coupling $\kappa_1$ which breaks  the Ising Z$_2\otimes$Z$_2$ symmetry down to Z$_2$. Most notably the degeneracy of the AFOy4 phase is lowered down to 2, and we expect AC and AO to become Ising type transitions. On the other hand the degeneracies of the three phases occurring near point $B$ are unchanged  and we expect the structure of the phase diagram in the vicinity of this MCP to remain unmodified up to some finite $\kappa_1$.  Recall that in the limit $\kappa_2\ll \kappa_1\ll \Delta$ the phase diagram is the one in Fig. \ref{fg:annni}.  We leave the evolution of the phase diagram as a function of $\kappa_1/ \kappa_2$ as an interesting open question.

\section{Discussion\label{sec:concl}}

In summary, we explored orbital ground states in the vicinity of a macroscopically degenerate classical transition. We have explicitly discussed  1D spin gapped systems where the energy scale of the orbital excitations is pushed down to low energies due to the competition between the Ising type interaction and the energy splitting induced by a lattice distortion (orbital field). Near the critical point, quantum fluctuations generated by spin-orbit coupling become important and turn the classical critical point into a multicritical point.

For the toy model in which a classical Ising orbital chain and a Haldane spin chain are coupled by weak relativistic spin-orbit interaction, we find that the effective orbital model has a multicritical point in the universality class of the quantum 1D ANNNI model. Besides the classical para-orbital and antiferro-orbital ordered phases, there appear near the multicritical point a phase with transverse orbital order and a critical orbital liquid phase. This Luttinger liquid phase arises without continuous orbital symmetry in the lattice model, in the region of the phase diagram between ordered phases with independent order parameters.

Motivated by experimental results for the compound CaV$_2$O$_4$, we also considered the effective orbital model  for a zigzag chain in the regime of dominant next-nearest-neighbor exchange coupling. We have found an exact solution for a U(1)$\otimes$U(1) symmetric model equivalent to hard-core fermions with  hopping along the legs of the zigzag chain and a nearest-neighbor exclusion constraint. This model  contains a doubly degenerate critical phase where the Luttinger liquid is confined to one of the legs.

More realistically, in the model where the U(1)$\otimes$U(1) symmetry is broken down to Z$_2$ (which stems from time reversal symmetry in the original spin-orbital model), we predict that the Luttinger liquid phase does not survive near the multicritical point, but is replaced by a gapped phase with coexisting orbital order parameters. The interpretation for this phase is that  the two legs have different polarization in the direction of the orbital field, as in the classical ordered phase, but at the same time the Z$_2$ symmetry is broken. This result should be relevant for CaV$_2$O$_4$ if the appropriate model parameters satisfy the hierarchy of energy scales assumed here, namely that the next-nearest-neighbor spin exchange is the dominant energy scale, much larger than the spin-orbit interaction. If the compound is also in the regime where the orbital field is comparable to the orbital Ising interaction, it would be interesting to search experimentally for the orbital multicritical point.

\section{Acknowledgements}
 We thank I. Affleck, F. C. Alcaraz, L. Fritz, M. Garst and A. Rosch for helpful discussions. This work was
supported by the A.V. Humboldt Foundation (E.S.).


\begin{thebibliography}{99}
\bibitem{tokura}Y. Tokura and N. Nagaosa, Science \textbf{288}, 462 (2000).
\bibitem{kugel}K.I. Kugel and D. Khomskii, Sov. Phys. Usp. \textbf{25}, 231 (1982).
\bibitem{khaliullin}G. Khaliullin, Prog. Theor. Phys. Suppl. \textbf{160}, 155 (2005).
\bibitem{itoi}C. Itoi, S. Qin, and I. Affleck, Phys. Rev. B \textbf{61}, 6747 (2000).
\bibitem{ulrich}C. Ulrich \textit{et al.}, Phys. Rev. Lett. \textbf{91}, 257202 (2003).
\bibitem{maekawa}G. Khaliullin and S. Maekawa, Phys. Rev. Lett. \textbf{85}, 3950 (2000).
\bibitem{hemberger}J. Hemberger \textit{et al.}, Phys. Rev. Lett. \textbf{91}, 066403 (2003).
\bibitem{niazi}A. Niazi  \textit{et al.}, Phys. Rev. B \textbf{79}, 104432 (2009).
\bibitem{pieper}O. Pieper \textit{et al.}, Phys. Rev. B  \textbf{79}, 180409(R) (2009).
\bibitem{chernperkins}G.-W. Chern and N. Perkins, Phys. Rev. B \textbf{80}, 220405(R) (2009).
\bibitem{Chern10}G.-W. Chern, N. Perkins and G. I. Japaridze, Phys. Rev. B \textbf{82}, 172408 (2010).
\bibitem{nersesyan}A. Nersesyan, G.-W. Chern, and N. B. Perkins, Phys. Rev. B \textbf{83}, 205132 (2011).
\bibitem{tsunetsugu}H. Tsunetsugu and Y. Motome, Phys. Rev. B \textbf{68}, 060405(R) (2003).
\bibitem{tchernyshyov}O. Tchernyshyov, Phys. Rev. Lett. \textbf{93}, 157206 (2004).
\bibitem{wheeler} E. M. Wheeler \textit{et al.}, Phys. Rev. B \textbf{82}, 140406(R) (2010).
\bibitem{haldane}F. D. M. Haldane, Phys. Lett. \textbf{93A}, 464 (1983); Phys. Rev. Lett. \textbf{50}, 1153 (1983).
\bibitem{whitehuse}S. R. White and D. A. Huse, Phys. Rev. B \textbf{48}, 3844 (1993).
\bibitem{BaxterBook}
R. J. Baxter, {\it Exactly solved models in statistical mechanics}, (Academic Press, London,
1982).
\bibitem{selke} W. Selke, Phys. Rep. \textbf{170}, 213 (1988).
\bibitem{Chakrabarti} B. K. Chakrabarti, A. Dutta and P. Sen  \textit{Quantum Ising Phases and Transitions in Transverse Ising Models}
(Berlin: Springer, 1996).
\bibitem{Villain} J. Villain and P. Bak, J. Phys. (Paris) \textbf{42}, 657 (1981).
\bibitem{Rujan} P. Ruj\'an, Phys. Rev. B \textbf{24}, 6620 (1981).
\bibitem{Allen} D.~Allen, P.~Azaria and P. Lecheminant, J. Phys. A: Math. Gen. \textbf{34}, L305 (2001).
\bibitem{feo} M. Beccaria, M. Campostrini and A. Feo, Phys. Rev. B \textbf{76}, 094410 (2007).
\bibitem{Santos} G. G\'omez-Santos, Phys. Rev. Lett. \textbf{70}, 3780 (1993).
\bibitem{Trippe} C. Trippe, F. G\"ohmann and A. Kl\"umper, J. Stat. Mech. P01021 (2010).
\bibitem{Giamarchi} T. Giamarchi, {\it Quantum Physics in One Dimension} (Oxford University Press, New York, 2004).
\bibitem{zuber} J. B. Zuber and C. Itzykson, Phys. Rev. D \textbf{15}, 2875 (1975).
\bibitem{AT} M. Kohmoto, M. d. Nijs and L. Kadanoff, Phys. Rev. B \textbf{24}, 5229 (1981).
\bibitem{sela2011} E. Sela, A. Altland, A. Rosch, arXiv:1103.4969  (unpublished).
\end{thebibliography}
\end{document}